\documentclass[letterpaper]{article} 
\usepackage{aaai25}  
\usepackage{times}  
\usepackage{helvet}  
\usepackage{courier}  
\usepackage[hyphens]{url}  
\usepackage{graphicx} 
\usepackage{natbib}
\usepackage{caption}  
\usepackage{amsfonts}
\usepackage{algorithm}
\usepackage{algorithmic}
\usepackage{bbding}
\usepackage{multirow}

\usepackage{newfloat}
\usepackage{listings}
\DeclareCaptionStyle{ruled}{labelfont=normalfont,labelsep=colon,strut=off} 
\floatstyle{ruled}
\newfloat{listing}{tb}{lst}{}
\floatname{listing}{Listing}

\title{GVMGen: A General Video-to-Music Generation Model \\with Hierarchical Attentions}
\author{
    Heda Zuo\textsuperscript{\rm 1},
    Weitao You\textsuperscript{\rm 1}\thanks{Corresponding author.},
    Junxian Wu\textsuperscript{\rm 1},
    Shihong Ren\textsuperscript{\rm 2},\\
    Pei Chen\textsuperscript{\rm 1},
    Mingxu Zhou\textsuperscript{\rm 1},
    Yujia Lu\textsuperscript{\rm 3},
    Lingyun Sun\textsuperscript{\rm 1}
}
\affiliations{
    \textsuperscript{\rm 1}College of Computer Science and Technology, Zhejiang University\\
    \textsuperscript{\rm 2}Shanghai Key Laboratory for Music Acoustics, Shanghai Conservatory of Music\\
    \textsuperscript{\rm 3}School of Design and Fashion, Zhejiang University of Science and Technology\\
   
    \{zuoheda, weitao\_you, wujunxian\}@zju.edu.cn, shren@shcmusic.edu.cn\\
    \{chenpei, zhoumingxu\}@zju.edu.cn, luyujia0426@zust.edu.cn, sunly@zju.edu.cn\\
   
}

\usepackage{bibentry}
\begin{document}

\maketitle

\begin{abstract}
Composing music for video is essential yet challenging, leading to a growing interest in automating music generation for video applications. Existing approaches often struggle to achieve robust music-video correspondence and generative diversity, primarily due to inadequate feature alignment methods and insufficient datasets. In this study, we present \textbf{G}eneral \textbf{V}ideo-to-\textbf{M}usic \textbf{Gen}eration model~(\textbf{GVMGen}), designed for generating high-related music to the video input. Our model employs hierarchical attentions to extract and align video features with music in both spatial and temporal dimensions, ensuring the preservation of pertinent features while minimizing redundancy. Remarkably, our method is versatile, capable of generating multi-style music from different video inputs, even in zero-shot scenarios. We also propose an evaluation model along with two novel objective metrics for assessing video-music alignment. Additionally, we have compiled a large-scale dataset comprising diverse types of video-music pairs. Experimental results demonstrate that GVMGen surpasses previous models in terms of music-video correspondence, generative diversity, and application universality.
\end{abstract}

\begin{links}
    \link{Code \& Cases}{https://chouliuzuo.github.io/GVMGen/}
\end{links}

\section{Introduction}

In videos, music plays a critical role in enhancing emotional resonance by aligning rhythm, style, and affectivity to achieve a high degree of correspondence. Traditionally, this task falls within the purview of professionals, while amateurs often find it challenging, characterized by time-consuming processes and potential copyright infringement. Consequently, the automatic generation of music based on video presents significant utility for both amateurs and industry professionals.

One of the core challenges in video background music generation is identifying the cross-modal relationships between visual and musical elements. Previous studies~\cite{di2021video,zhu2023discrete,yu2023longtermrhythmicvideosoundtracker,zhu2022quantized} define rule-based connections between specific variables, such as motion speed and color. However, these variables are only significant for certain types of videos which are less important for video blogs or documentaries~\cite{Corner_2002}. Many other variables, such as shots and compositions, are overlooked despite the high relevance, particularly in movies. Studies like \cite{hussain2023m,tang2024any} rely on Large Language Models~(LLMs) as a bridge. However, LLMs often summarize the video and music into some static styles while ignoring details like emotions and the exact physical variables with temporal changes. Such explicit feature alignment may ignore high-related information which cannot be computed or described, while considering redundant unrelated features, thus limiting the depth and coherence of the music-video correspondence. 

Moreover, music features derived from variable or language transformations are neither diverse nor detailed enough to guide vivid and artistic music generation. Many of these approaches are even constrained to MIDI music~\cite{su2020audeo,gan2020foley}, which simply encodes each musical note as a numeric symbol. Consequently, the generated music from these models tends to be monotonic and lacks rich diversity and universality.

In addition, inadequate evaluation metrics and datasets further constrain the effectiveness of video-to-music generation models. Most existing works assess music-video correspondence primarily through subjective evaluation~\cite{ji2020comprehensivesurveydeepmusic,D2022It}, which is costly and often biased, failing to guide the model training correctly. Existing datasets~\cite{zhuo2023video, kang2024video2music} primarily consist of music videos (MVs) with music in MIDI format. These datasets exhibit low diversity and weak music-video correspondence, thereby limiting the efficacy of models trained on them.

In this paper, we propose a \textbf{G}eneral \textbf{V}ideo-to-\textbf{M}usic \textbf{Gen}eration model~(\textbf{GVMGen}), which can generate high-related music in various styles for different types of video. Unlike previous models, GVMGen refrains from explicitly defining variable relationships or relying on language transformation between visual and musical features. Instead, we extract hidden visual features through spatial self-attention and transform them into musical features through both spatial and temporal cross-attention. By adopting an implicit attention mechanism for feature transformation, GVMGen preserves the most relevant features in alignment, thereby enhancing the music-video correspondence. Moreover, implicit feature extraction and alignment are well-suited for different styles of video and music, enabling GVMGen to be a general model that performs well even in zero-shot cases.

Moreover, we propose an evaluation model with two novel objective metrics assessing both global cross-modal relevance and local temporal alignment. We also collect a large-scale video-music dataset that encompasses a diverse range of styles, including movies, video blogs (vlogs), and more, rather than relying solely on MVs in MIDI format. Furthermore, our dataset includes a significant portion of Chinese traditional music performed on over ten types of instruments. Chinese traditional music emphasizes diverse fingering techniques and complex timbres, which cannot be represented by MIDI files and simple musical variables. This inclusion introduces a higher level of difficulty but also enhances the diversity in music generation.

\par Experimental results reveal that GVMGen exhibits robust performance, particularly in terms of music-video correspondence and music richness. Both the generative similarity with ground truth and the quality of the music are improved simultaneously. GVMGen can generate multi-track waveform music over MIDI in both Chinese and Western styles, marking a pioneering advancement in the richness and completeness of music generation. Furthermore, GVMGen demonstrates remarkable universality, enabling high-quality music generation even in zero-shot scenarios.
\par In summary, our main contributions can be written as: 
\begin{itemize}
    \item We propose GVMGen, a general video-to-music generation model based on hierarchical attentions, capable of generating diverse genres of music highly related to different styles of videos.
    \item We propose an evaluation model with objective metrics for local and global music-video correspondence evaluation, and also collect a large-scale video-music dataset encompassing multiple styles of both video and music.
    \item We conduct extensive experiments which show that our model can outperform state-of-the art models significantly in terms of video-music correspondence, music diversity and application universality.
\end{itemize}

\section{Related Work}
\textbf{Music Generation} can be divided into symbolic music generation and waveform music generation. For symbolic music generation, MIDI-VAE~\cite{brunner2018midivae} and MusicVAE~\cite{roberts2018musicvae} adopt variational autoencoder while Music Transformer~\cite{huang2018music} and MuseGAN~\cite{Dong_Hsiao_Yang_Yang_2018} use attention-based sequence generation and adversarial generation techniques. These models can only generate MIDI music. For waveform music generation from text description, Riffusion~\cite{Forsgren_Martiros_2022} employs the pretrained Stable Diffusion model to transform text-to-music process into a text-to-spectrogram task, thereby enabling the generation of music. As for large music generation models, Google proposes MusicLM~\cite{agostinelli2023musiclm} based on Mulan~\cite{huang2022mulan} and SoundStream~\cite{zeghidour2021soundstream}. MusicGen~\cite{copet2024simple} uses T5~\cite{JMLR:v21:20-074} as a text encoder and utilizes quantized music codes from Encodec~\cite{defossez2022highfi} for generation, while Stable Audio~\cite{evans2024fast} adds a diffusion UNet into MusicGen. These models provide a foundational structure for works related to music generation.
\\\textbf{Video Background Music Generation} is proposed by~\cite{di2021video} which uses rule-based computation to predict music features. V-MusProd~\cite{zhuo2023video}, V2Meow~\cite{su2023v2meow} and Video2Music~\cite{kang2024video2music} rely more on deep neural network to extract visual features for music generation and propose several evaluation metrics. With the help of LLMs, CoDi~\cite{tang2024any} and  M$^2$UGen~\cite{hussain2023m} use LLMs as a bridge to achieve cross-modal generation through the help of semantic description. \cite{Li_2024_CVPR} proposes a diffusion generation model with segment-aware cross-attention. However, the music-video correspondence is still constrained due to the limited explicit features or insufficient alignment. Therefore, we adopt hierarchical attentions to do cross-modal feature alignment in both spatial and temporal aspects which is more accurate and universal.
\\\textbf{Video-music Dataset} is a kind of multi-modal dataset which is dedicated to video background music generation task. AIST++~\cite{Li_2021_ICCV} and TikTok~\cite{zhu2022quantized} datasets contain dance videos, accompanied with music and visual motion information. However, these datasets are limited by its style diversity and total duration. Then, SymMV~\cite{zhuo2023video} and MuVi-Sync~\cite{kang2024video2music} datasets provide over 50 hours of music videos with music feature annotations, primarily suitable for symbolic music generation. Recently, with the strong comprehension and language processing ability of LLMs, M$^2$UGen proposes a systematic approach for generating datasets through music oriented instructions.~\cite{hussain2023m}. Since these dataset are not logically suitable and diverse enough, we collect a large-scale dataset encompassing both Chinese traditional music and western music with various types like movies, vlogs and so on.
\begin{figure*}
    \includegraphics[width=\textwidth]{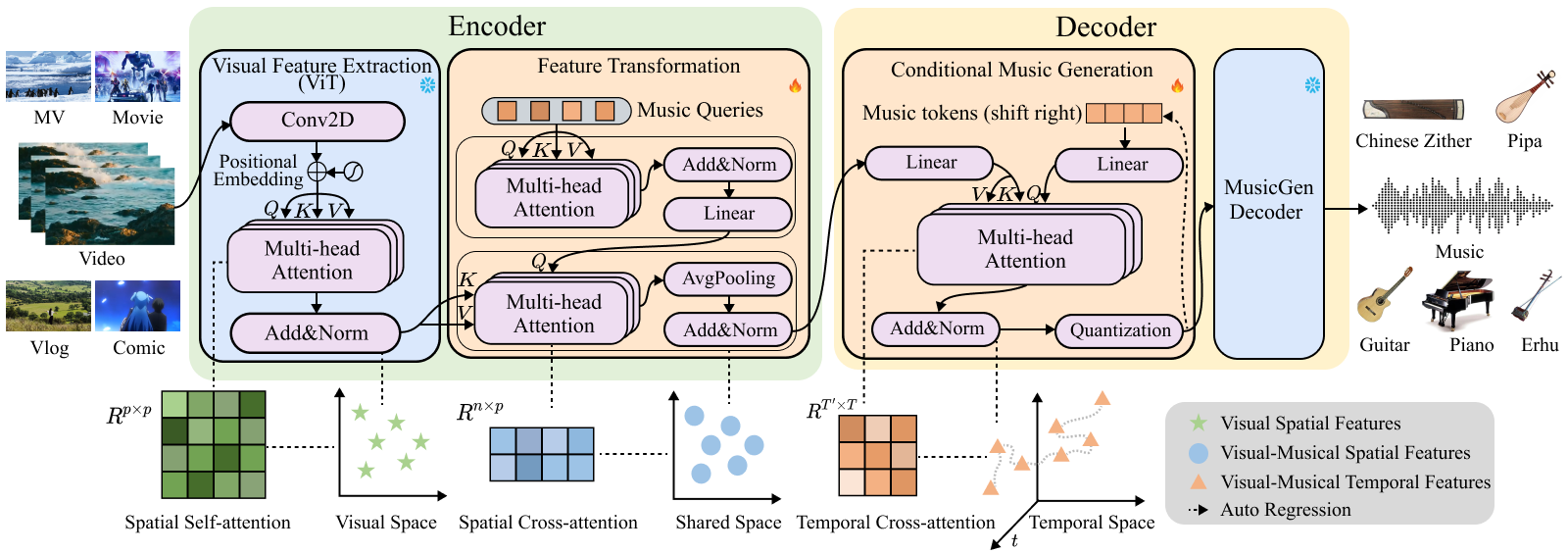}
    \caption{General Video-to-Music Generation (GVMGen) model with encoder-decoder struture. The model consists of: (1) Visual feature extraction module with spatial self-attention; (2) Feature transformation module with spatial cross-attention; (3) Conditional Music generation module with temporal cross-attention.}
    \label{fig:2}
\end{figure*}
\section{Method}
In this section, we first define the video background music generation problem, then present GVMGen model in details, together with theoretical derivation and analysis.
\subsection{Problem Definition}
In video background music generation, suppose we are given a dataset with $N$ samples of video and music pairs. We use $\mathbf{V}\in \mathbb{R}^{t\times f\times H\times W\times C}$ to denote each piece of video, where $t$, $f$, $H$, $W$, $C$ stands for duration, video frame rate and height, width, number of channels of image separately. Music is denoted as quantized codes $\mathbf{M}\in \mathbb{R}^{t\times f'\times K}$. $f'$ stands for music code sample rate and $K$ stands for the number of codebooks. The training tuples $(\mathbf{V},\mathbf{M})^{N_{train}}$ contain $N_{train}$ instances while other $N-N_{train}$ samples form the test set. The goal of video background music generation is to accurate generate $M$ for the test set which is more similar to the original high-related music.
\subsection{General Video-to-Music Generation Model}
As shown in Figure~\ref{fig:2}, GVMGen is an end-to-end video-to-music model leveraging hierarchical attentions. Initially, visual features are extracted from the video through visual feature extraction module using spatial self-attention. Subsequently, a feature transformation module equipped with trainable music queries filters the extracted visual features to retain those relevant to music through spatial cross-attention. Finally, the conditional music generation module is facilitated by temporal cross-attention. All features are processed as deep hidden features, which is fairly appropriate for different styles of video and music. As a result, GVMGen is able to focus on the most related feature with minimal information loss, thereby generating diverse music with high relevance to the video. 
\par \textbf{Attention mechanism} is proposed by ~\cite{vaswani2017attention} as a main component of FFTs, which is a function mapping a query and a set of key-value pairs to an output. In detail, the weight is calculated by a compatibility function~(usually dot product) of the query with the corresponding key. After normalization, the weights are assigned to each value. We set $Q$, $K$ and $V$ stand for query, key and value separately while $D_k$ stands for the dimension of key, the attention mechanism $f(\cdot)$ can be written as:
\begin{equation}
    f(Q,K,V) = softmax(\frac{QK^T}{\sqrt{D_k}})V
\end{equation}
Intuitively speaking, attention mechanism calculates the relevance between key-value pairs and queries. A higher weight indicates that the pair is more relevant to the query.
\par \textbf{Visual feature extraction module.} Video cannot be directly transferred into music as they belong to different latent spaces. Therefore, the cross-modal relationship must be built on related features.

Since deep features can preserve a greater amount of information than several variables, in GVMGen, we use a pretrained VIT-L/14@336px~\cite{dosovitskiy2021image} with spatial self-attention to extract deep visual features. ViT is the image encoder of CLIP~\cite{radford2021learningtransferablevisualmodels}, which splits an image into $p$ patches and transforms them into embeddings. For an image embedding $x$, spatial self-attention $f(x,x,x)$ is used to derive the importance of each patch. Assume $w_{ij}$ is each element of the matrix $a(Q,K) = \frac{QK^T}{\sqrt{D_k}}$, the extracted feature $z$ can be calculated as:
\begin{equation}
    z_i = \Sigma^k_{j=1} w_{ij}x_i, \Sigma w_{ij} = 1
\end{equation}
which is a linear transformation from the original embeddings. This feature extraction method can derive deep relationship while preserve the original information as well.

In this approach, we treat the video as a sequence of images and focus on extracting the inner deep features of these images. Our goal is to preserve both the spatial and temporal features for transformation in the subsequent modules, as music also encompasses spatial and temporal dimensions. If other video models like~\cite{arnab2021vivit,xu2021videoclip} are adopted in visual feature extraction, the temporal information may be lost, which would hinder the music generation process in terms of temporal alignment. This loss in temporal alignment would consequently reduce the correspondence between the music and the video. The results of our ablation study further support this assertion.
\par \textbf{Feature Transformation module.} This is one of the most critical parts in video-to-music generation, as it is essential to establish the cross-modal relationship in the shared (visual-musical) space. We do not rely on either mathematical relationship definition or language descriptions for transformation like previous works, since the former is limited by incomplete and inaccurate variables, while the latter tends to lose temporal information and introduces redundant information from an extra modality.

In GVMGen, we propose spatial cross-attention to build the gap between visual and musical features inspired by~\cite{li2022blip}. Firstly, we define trainable music queries $q$. The queries interact with each other through self-attention, and interact with the extracted visual features $z$ through cross-attention. Since the attention $a(q,z)$ identifies the relevance between visual and musical spaces, it establishes the cross-modal relationship which can be viewed as a shared space. Through $z'=f(q,z,z)$, the visual feature is transformed into this space like a projection, and the transformed features $z'$ can represent the cross-modal features. Consequently, the most relevant features are preserved while the redundant unrelated features are filtered out after feature transformation.

Unlike previous methods that utilize variables or language, which are stable and may only be effective for certain types of video, the transformation by cross-attention focuses on the relevant features of each distinct video. This approach leverages a shared space informed by both music queries and the visual input itself. Consequently, the transformation output, which subsequently governs music generation, is diverse and contingent upon the video input. The number of queries plays a crucial role in determining the size of the shared space and the transformed features. Our experimental results indicate that utilizing 16 music queries yields optimal performance in filtering and transforming features.

\par \textbf{Conditional Music Generation module.} Finally, the extracted features are essential for reconstructing the music. Since both video and music encompass spatial and temporal information, in GVMGen, we design a temporal cross-attention to guide music generation with temporal alignment. This module functions similarly to a decoder-only transformer, where the query vector is derived from the real music embedding $m$~(shifted right). The temporal cross-attention operates as $m'=f(m,z',z')$, where the attention weight $a(m,H)$ is an $R^{T'\times T}$ matrix. Here, $T = t\times f$ and $T'=t\times f'$ represent the duration, and the attention aligns cross-modal feature with music embedding on a temporal basis, thereby ensuring temporal visual-musical correspondence. Additionally, the attention weight elucidates the relationships among contextual features, thereby reinforcing global dependency and the integrity of the generated music.

After temporal cross-attention, a pretrained MusicGen~\cite{copet2024simple} decoder is used to decode the music embedding into music in audio format. The MusicGen utilizes Encodec~\cite{defossez2022highfi} as its decoder, which applies Residual Vector Quantization~(RVQ) to compress audio flow into discrete tokens. This approach reduces space complexity and possesses the capability to generate diverse music due to extensive data training. Consequently, we incorporate it as part of the GVMGen decoder. Additionally, to enhance generative diversity and universality, we have curated a more vivid video-music dataset for training.

\begin{table*}
\footnotesize
    \centering
    \begin{tabular}{c|ccccc|ccc|c}
         Dataset& MV & Movie & Vlog & Comic &Documentary & Western music& Chinese music&Ensemble &Length\\
         \hline
         TikTok&\XSolidBrush&\XSolidBrush&\Checkmark&\XSolidBrush&\XSolidBrush&\Checkmark&\XSolidBrush&\XSolidBrush&1.5h\\
         AIST++&\XSolidBrush&\XSolidBrush&\Checkmark&\XSolidBrush&\XSolidBrush&\Checkmark&\XSolidBrush&\XSolidBrush&5.2h\\
         SymMV&\Checkmark&\XSolidBrush&\XSolidBrush&\XSolidBrush&\XSolidBrush&\Checkmark&\XSolidBrush&\XSolidBrush&78.9h\\
         MuVi-Sync&\Checkmark&\XSolidBrush&\XSolidBrush&\XSolidBrush&\XSolidBrush&\Checkmark&\XSolidBrush&\XSolidBrush&54.6h\\
         \hline
         Ours (CTM)&\Checkmark&\XSolidBrush&\XSolidBrush&\XSolidBrush&\XSolidBrush&\XSolidBrush&\Checkmark&\XSolidBrush&89.5h\\
         Ours&\Checkmark&\Checkmark&\Checkmark&\Checkmark&\Checkmark&\Checkmark&\Checkmark&\Checkmark&147h\\
         
    \end{tabular}
    \caption{Comparison between different video-music datasets in which `CTM' stands for Chinese traditional music part. Our dataset is the first dataset that include both Chinese traditional music and western music.}
    \label{tab:my_label1}
\end{table*}
\begin{figure}
    \centering
    \includegraphics[width=\linewidth]{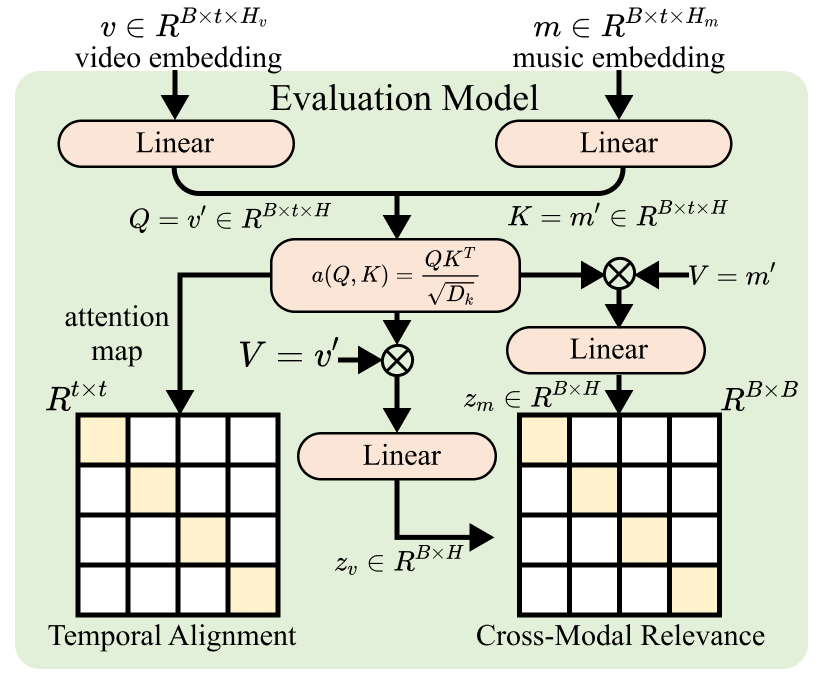}
    \caption{Evaluation model with both Temporal Alignment (TA) and Cross-Modal Relevance (CMR), where $z_v$ and $z_m$ represent video features and music features.}
    \label{fig:enter-label}
\end{figure}
\subsection{Evaluation Model}
Previous objective metrics such as Fréchet Audio Distance (FAD) and Kullback-Leibler Divergence (KLD), primarily focus on evaluating the similarity between the generated music and the ground-truth. However, these metrics do not account for the correspondence between video and music. 

Inspired by \cite{D2022It}, we propose an evaluation model for music in audio format with both global and local (temporal) estimation. For music-video pairs of batch size $B$, we transform the embeddings of hidden dimension $H_v$ and $H_m$ into a unified hidden size $H$. The cross-attention matrix $a(v,m)\in R^{t\times t}$ focuses on visual-musical relationship at each moment, effectively providing a temporal alignment for local evaluation. The hidden video and music features are derived from the cross-attention matrix with different values like $z_v = f(v,m,v)$ and $z_m=f(v,m,m)$. After linear layer to summarize the features for each video and music, the cross-modal relevance can be considered as a global evaluation metric inspired by~\cite{radford2021learningtransferablevisualmodels}.

For training, the temporal alignment employs MSELoss $\mathcal{L}_1$ to maximize diagonal attention since the local visual-musical correspondence should be strongest, while InfoNCE Loss $\mathcal{L}_2$ is employed for cross-modal relevance like:
\begin{equation}
    \mathcal{L}_1 = \frac{1}{t}\sum^t\sqrt{(I-diag(a(v,m)))^2}
\end{equation}
\begin{equation}
    \mathcal{L}_2 = \frac{1}{2}(\mathcal{L}_{m\rightarrow v} + \mathcal{L}_{v\rightarrow m})
\end{equation}
\begin{equation}
    \mathcal{L}_{m\rightarrow v} = -\sum_{i}^N\Big[log\frac{exp[s(f_m^i,f_v^i)/\tau]}{\sum_{j}^N exp[s(f_m^i,f_v^j)/\tau]}\Big]
\end{equation}
where $I$ stands for identity matrix, $s(f_m,f_v)=\frac{f_m^Tf_v}{\Vert f_m\Vert\cdot\Vert f_v\Vert}$. The temperature parameter $\tau$ is set to 0.07. During evaluation, the temporal alignment metric and cross-modal relevance metric are derived from the average of $diag(\cdot)$.

For efficiency, the evaluation model reach an average loss and accuracy of 0.003 and 99.4\% on the test set. We also find 20 expert users to score this metric on 30 video-music pairs, and the error rate between the average score and the model score is only 3.75\%.
\subsection{Training Process}
\par In the visual feature extraction module, each video $\mathbf{V}\in \mathbb{R}^{t\times f\times H\times W\times C}$ is considered as a sequence of images $p_{i}\in \mathbb{R}^{H\times W\times C}$. Each image will be divided into several patches represented by $x_{i}\in \mathbb{R}^{h\times w\times D}$, where $h=H/s, w=W/s$ and $s$ stands for patch size. After spatial self-attention, the hidden features can be represented by $F^p_{i}\in\mathbb{R}^{p\times D}$, where $p=h\times w+1$ in which the addition value stands for special class token~(cls).
\par In the feature transformation module, we create trainable music-related queries $q\in \mathbb{R}^{n\times D}$. Both self-attention and cross-attention are employed on
trainable queries. Here $n$ and $D$ stands for number of queries and hidden dimension, while $K$ and $V$ are $z^p_{i}$ in cross-attention.
\par After layers of cross-attention, we get music-relevant attention $A_{i}\in \mathbb{R}^{n\times D}$ for each image. We add an average pooling for attentions with different queries:
\begin{equation}
    z'^p_{i} = \frac{\sum_{j=1}^n A^j_{i}}{n}, A^j_{i}\in \mathbb{R}^{1\times D}
\end{equation}
\par Then the cross-modal features can be represented by $z'^p_{i}\in\mathbb{R}^{1\times D}$. For video, $z^v\in \mathbb{R}^{t\times f\times D}$ is stacked by a sequence of $z'^p_{i}$ for $i\in [0, t \times f]$.
\par In the conditional music generation module, the cross-modal features $z^v$ are then sent into decoder. After linear projection, music embeddings are set as queries, while cross-modal feature $z^v$ are set as keys and values. In training, music embeddings $m\in \mathbb{R}^{t\times f'\times K\times H}$ are from ground truth music tokens $\mathbf{M}\in \mathbb{R}^{t\times f'\times K}$~(shift right). $f'$ stands for music token sample rate and $K$ stands for the number of codebooks. When inference, music embeddings are initialized to a start token and will be iterated to predicted music embeddings auto-regressively. The predicted music embeddings can be written as $m'$ and will be quantized into music tokens $\mathbf{M}'$ according to codebooks. The music tokens will be decoded as music in audio format finally.
\par For training loss, we adopt average cross entropy loss of each codebook $B_j$ to compare the predicted music features $F'^m$ and the ground truth music tokens $M$:
\begin{equation}
    \mathcal{L} = -\frac{1}{NK} \sum_{j=1}^{K} \sum_{i=1}^{N} B_j \mathbf{M}_i \log({z^m_i}')
\end{equation}
\subsection{Dataset}
To enhance the generative diversity and universality, we collect a large-scale video-music dataset encompassing various types of videos and music. Existing datasets primarily consist of music videos~(MVs) ~\cite{zhuo2023video,kang2024video2music} with MIDI music. However, for MVs, videos are typically produced after the music has been composed, which is logically contrary to the task of generating background music for a given video. Such datasets exhibit low diversity and weak music-video correspondence, thereby limiting the efficacy of models trained on them.
\par To address these limitations, our dataset comprises movies, vlogs, comics, and documentaries where the background music is specifically tailored for the video content. For music, our datasets include a substantial amount of Chinese traditional music as well as ensembles featuring both Chinese and Western instruments. Chinese traditional music emphasizes intricate melodies and rhythms~\cite{liu_aesthetic_1985}, along with variations in the playing techniques that cannot be adequately represented in MIDI format. As shown in Table~\ref{tab:my_label1}, our dataset spans a wide range of topics and includes various types of background music. This diversity is crucial for developing robust and versatile models capable of generating appropriate music for different video genres and styles.
\par For collection, we sourced our dataset from free public platforms (Bilibili and Youtube). We selected clips featuring solely music and excluded those video frames that contained extensive superimposed text or captions. After manual filtering, the dataset is preprocessed by clipping for training. The total durations are 89.5 hours for MVs, 42.1 hours for documentaries, 9.9 hours for vlogs, and 5.5 hours for other types.

 \begin{table*}
 \footnotesize
    \centering
    \begin{tabular}{c|ccc|ccc}
         \multirow{2}{*}{Model}&\multicolumn{3}{c|}{Non expert}&\multicolumn{3}{c}{Expert}\\
         \cline{2-7}
         \multicolumn{1}{c|}{}&OMQ$\uparrow$&MVC$\uparrow$&MR$\uparrow$&OMQ$\uparrow$&MVC$\uparrow$&MR$\uparrow$\\
         \hline
         CMT&3.52$\pm$0.11&2.85$\pm$0.14&3.20$\pm$0.40&2.35$\pm$0.15&2.23$\pm$0.19&2.40$\pm$0.95\\
         V2M&3.64$\pm$0.10&2.53$\pm$0.13&1.08$\pm$0.44&2.18$\pm$0.15&1.66$\pm$0.17&1.00$\pm$0.72\\
         M$^2$UGen&3.83$\pm$0.11&2.54$\pm$0.14&5.10$\pm$0.56&3.49$\pm$0.17&2.31$\pm$0.20&3.35$\pm$1.00\\
         \hline  
         Our (S,CTM)&3.29$\pm$0.12&2.72$\pm$0.17&-&2.76$\pm$0.17&2.19$\pm$0.24&-\\
         Our (S,FTM,CTM)&4.41$\pm$0.11&2.83$\pm$0.15&-&3.25$\pm$0.18&2.41$\pm$0.23&-\\
         Our (L,FTM,CTM)&\textbf{4.57$\pm$0.11}&3.02$\pm$0.17&-&3.59$\pm$0.17&2.55$\pm$0.25&-\\Our (L,FTM,All)&4.55$\pm$0.10&\textbf{4.62$\pm$0.14}&\textbf{5.98$\pm$0.40}&\textbf{4.25$\pm$0.16}&\textbf{4.91$\pm$0.20}&\textbf{6.20$\pm$0.45}\\

    \end{tabular}
    \caption{Subjective evaluation with 95\% confidence interval. Here 'S' and 'L' indicates small and large~(24 temporal cross-attention layers with 493M parameters and 48 with 1.9B), 'FTM' stands for using the feature transformation module and 'CTM' means the model is only trained on our CTM dataset while 'All' means our whole dataset.}
    \label{main}
\end{table*}

\begin{table*}
\footnotesize
    \centering\resizebox{\textwidth}{!}{
    \begin{tabular}{c|ccccc|ccccc|ccccc}
         \multirow{2}{*}{Model}&\multicolumn{5}{c|}{SymMV~(MVs)}&\multicolumn{5}{c|}{MuVi-Sync~(MVs)}&\multicolumn{5}{c}{Other~(movies and comics)}\\
         \cline{2-16}
         \multicolumn{1}{c|}{}&KLD$\downarrow$&CMR$\uparrow$&TA$\uparrow$&OMQ$\uparrow$&MVC$\uparrow$&KLD&CMR&TA&OMQ&MVC&KLD&CMR&TA&OMQ&MVC\\
         \hline
         CMT&2.65&0.67&0.85&3.34&2.81&1.75&0.34&0.69&3.27&2.81&0.69&0.75&0.75&3.05&2.69\\
         V2M&2.73&0.50&\textbf{0.91}&3.00&1.91&1.89&0.21&\textbf{0.78}&2.90&1.81&0.67&0.72&\textbf{0.77}&3.14&2.47\\
         M&2.61&\textbf{0.93}&0.89&3.64&2.46&1.59&\textbf{0.69}&0.68&3.66&2.59&0.84&0.96&0.75&3.53&2.98\\
         \hline
         Our&\textbf{2.20}&\textbf{0.93}&0.88&\textbf{4.43}&\textbf{4.08}&\textbf{1.37}&0.68&0.70&\textbf{4.00}&\textbf{4.36}&\textbf{0.45}&\textbf{0.97}&0.69&\textbf{4.43}&\textbf{5.16}
         
    \end{tabular}}
    \caption{Universality evaluation on other video-music dataset, where M stands for M$^2$UGen.}
    \label{main2}
\end{table*}
\section{Experiment}
\subsection{Implementation Details} We adopt VIT-L/14@336px with 24 self-attention layers. In feature transformation module, we employ 16 queries, 6 self-attention layers and 3 cross-attention layers. And the temporal cross-attention is with 48 transformer layers of 1536 as the hidden size while the MusicGen decoder is with 4 codebooks of 2048 tokens. We use Adam optimizer with learning rate of 1e-5, weight decay of 0.01, batch size of 6 and video frame rate of 1 per second. A cosine learning rate schedule with 4000 warmup steps and top-k sampling with keeping the top 250 tokens are employed. We use 30-second clips with the music sampling rate of 32kHz. The ratio of training and test set is 0.85:0.15. The training lasts for 150 epochs with 188 hours on NVIDIA A100~(single card).
\subsection{Metrics}
We evaluate the models with both objective and subjective metrics. For objective metrics, we adopt FAD, KLD~\cite{gemmeke2017audio} which are commonly utilized to evaluate the relevance between original and generated music. Following the evaluation model, We further use cross-modal relevance (CMR) and temporal alignment (TA) to evaluate the music-video correspondence both in global and temporal aspect. For FAD and KLD, a lower score indicates a more similar music, while for CMR and TA, a higher score indicates the music is more related and well-aligned.

\par We also invited 20 non-expert listeners and 10 with professional music knowledge for subjective evaluation. We require listeners to evaluate music samples in three aspects: (1) overall music quality~(OMQ), which evaluates the rationality and quality of the music; (2) music-video correspondence~(MVC) which evaluates the semantic, emotional and rhythmic consistency between the music and video; (3) music richness~(MR) which evaluate the generative diversity. OMQ and MVC are scored for a single music sample while MR is for a series of samples generated by each model. We use 32 samples to evaluate OMQ and MVC by average score while a set of 10 samples of each to evaluate MR. All these metrics are based on 7-point scale.
\subsection{Comparison Models}
For comparison, we choose CMT~\cite{di2021video}, V2M~\cite{kang2024video2music}, M$^2$UGen~\cite{hussain2023m}, NExT-GPT~\cite{wu2024nextgptanytoanymultimodalllm} and CoDi~\cite{tang2024any} as baseline models. CMT and V2M are based on Music Transformer which can only generate MIDI music, while other three leverage LLMs to build a bridge for multi-modal music generation and understanding. The evaluation set comprises of data from our test set, SymMV\cite{zhuo2023video}, MuVi-Sync\cite{kang2024video2music} and other random data. In objective evaluation, we select 10 pieces of data for each kind resulting in a total of 40 pieces of data in the evaluation set. Each piece of data has a duration of 15 seconds, providing a sufficient length to capture meaningful visual and musical elements while keeping the evaluation process manageable.
\subsection{Experimental Results}
\begin{table}[]
\small
    \centering\scalebox{0.95}{
    \begin{tabular}{c|cccc}
         Model &FAD$_{vgg}$$\downarrow$&KLD$\downarrow$&CMR$\uparrow$&TA$\uparrow$\\
         \hline
         CMT&12.29$\pm$1.12&1.98$\pm$0.70&0.65$\pm$0.10&0.78$\pm$0.09\\
         V2M&21.50$\pm$1.32&2.03$\pm$0.69&0.51$\pm$0.10&0.82$\pm$0.09\\
         M&11.23$\pm$2.63&1.98$\pm$0.68&0.88$\pm$0.07&0.76$\pm$0.10\\
         N&23.21$\pm$1.08&4.80$\pm$0.31&0.44$\pm$0.11&\textbf{0.83$\pm$0.07}\\
         CoDi&17.40$\pm$2.59&2.96$\pm$0.79&0.69$\pm$0.08&0.73$\pm$0.11\\
         \hline       
         Our&\textbf{7.45$\pm$1.62}&\textbf{1.40$\pm$0.55}&\textbf{0.88$\pm$0.06}&0.79$\pm$0.09
    \end{tabular}}
    \caption{Objective evaluation with 95\% confidence interval, where M stands for M$^2$UGen and N stands for NExT-GPT.}
    \label{tab1}
\end{table}
\begin{figure*}
    \includegraphics[width=\textwidth]{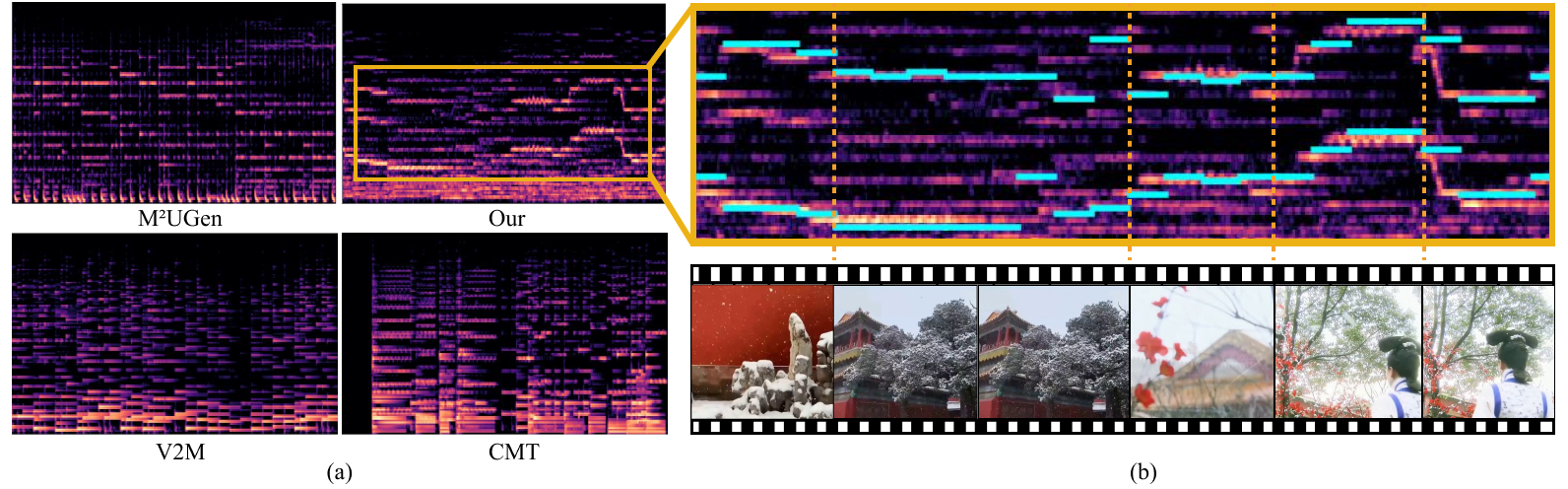}
    \caption{Mel-spectrums of generated music by models according to the same video input in (b), (b) illustrates the pitch contours and alignment of the music generated by our model with the video input.}
    \label{res}
\end{figure*}
\textbf{Performance Comparison.} Table~\ref{tab1} presents the performance of objective metrics on the evaluation set. It is evident that GVMGen produces music most closely resembling the ground truth while maintaining a high relevance to the corresponding video. Given that NExT-GPT and CoDi underperform in at least one metric and their generated samples are more like audio rather than music, they are excluded from further subjective evaluation.

Table~\ref{main} illustrates the performance of subjective metrics as evaluated by both experts and non-experts. The results indicate that: (1) Our model outperforms others across all metrics, with significant improvements particularly in music-video correspondence and music richness. This suggests that our model is capable of generating a diverse styles of music that are high-related to the video input; (2) Even when GVMGen is trained solely on our Chinese Traditional Music dataset, its performance in music quality and music-video correspondence remains comparable to, or even surpasses, that of other models. This demonstrates the strong universality and transferability of our model.

\textbf{Visualization.} As shown in Fig.~\ref{res} (a), with the same video input, only our model can generate music with two distinct thematic lines, played by the erhu and piano, whereas other models fail to produce a clear melody. Fig.~\ref{res} (b) demonstrates that the music generated by our model fluctuates synchronously with the video shots, descending during background scenes and reaching a climax when the protagonist appears. This indicates that our model can generate music with precise temporal alignment to the video.

\textbf{Universality Study.} Table~\ref{main2} shows the detailed performance on other datasets. GVMGen consistently outperforms other models, even on their datasets, whether in terms of similarity to the ground truth, generative quality, or music-video correspondence. These datasets are entirely different from our training set, indicating that GVMGen can be effectively applied to various types of video inputs, even in zero-shot scenarios.
\begin{table}[]
\footnotesize
    \centering
    \begin{tabular}{c|cccc}
         Model  & KLD$\downarrow$& FAD$_{vgg}$$\downarrow$& CMR$\uparrow$ & TA$\uparrow$ \\
         \hline
         Q4 avg&1.20&3.25&0.97&0.76\\
         Q8 avg&1.22&3.69&0.96&0.77\\
         Q32 avg&1.26&3.04&\textbf{0.99}&0.77\\
         \hline
         Q16 &1.14&2.73&0.95&0.78\\
         Q16 sum &\underline{1.10}&\textbf{2.55}&0.94&\underline{0.79}\\
         \hline
         Q16 w. ViViT&1.25&3.07&0.92&0.77\\
         Q16 w/o. TCA&1.75&16.71&0.35&0.71\\
         \hline
         \textbf{Our ~(Q16 avg)}&\textbf{1.07}&\underline{2.63}&\underline{0.98}&\textbf{0.80}\\
    \end{tabular}
    \caption{Objective ablation Study, where Q represents number of queries and TCA indicates temporal cross-attention.}
    \label{tab:my_label}
\end{table}
\subsection{Ablation Study}
To further study the effectiveness of each component in our model, we adopt additional ablation study with spatial cross-attention, temporal cross-attention. The objective ablations are performed using 1628 samples of 30 seconds in our test set while subjective evaluation are adopted with the main experiment using 32 samples of 15 seconds.
\par \textbf{Spatial Self Attention.} The model cannot work if we drop the module. And as shown in Table~\ref{tab:my_label}, if we change the visual feature extraction model from ViT into ViViT, whether the generative similarity or the music-video correspondence will drop significantly, confirming that it cause a information loss in temporal features as mentioned before.
\par \textbf{Spatial Cross Attention.} As shown in Table~\ref{main}, without the spatial cross attention, the performance of Our (S,CTM) falls sharply compared with Our (S,FTM,CTM). It indicates that if we remove the spatial cross-attention, the visual features cannot be transformed into shared space, which causes a lower generative quality and music-video correspondence. Moreover, we use different number of queries to test the most appropriate one with the average value, sum value or preserving the original cross-modal features. Results in Table~\ref{tab:my_label} shows that the 16-query model with average values generally performs best because it can focus on more effective features while preserve less redundant features.
\par \textbf{Temporal Cross Attention.} Table ~\ref{tab:my_label} shows that the generated music cannot be similar to ground truth or be related to video without temporal cross-attention. And in Table~\ref{main}, the adoption of more layers yields significant improvements across almost all metrics. This enhancement can be attributed to the increased capacity and complexity, allowing it to capture temporal alignment more accurately and keep the music of high relevance to the video. 

\section{Conclusion}
In this paper, we present GVMGen, capable of producing diverse music in audio format that is highly related to various types of video inputs. We leverage hierarchical attentions, including spatial self-attention, spatial cross-attention and temporal cross-attention, to extract and align the hidden features. The hierarchical attentions can preserve the most important features with minimal information loss. Moreover, we propose an evaluation model with two novel objective metrics to evaluate global and local music-video correspondence. We also collect a large-scale dataset including MVs, movies and vlogs, featuring both Chinese and western background music. Experimental results demonstrate that our model excels in the correspondence, diversity, and universality of video background music generation. We will improve the robustness and personalized generation in future.
\section{Acknowledgements}
This work is supported by the National Key Research and Development Program of China (2023YFF0904900), and the National Natural Science Foundation of China (No. 62272409). 
\bibliography{GVMGen}

\begin{thebibliography}{37}
\providecommand{\natexlab}[1]{#1}

\bibitem[{Agostinelli et~al.(2023)Agostinelli, Denk, Borsos, Engel, Verzetti, Caillon, Huang, Jansen, Roberts, Tagliasacchi et~al.}]{agostinelli2023musiclm}
Agostinelli, A.; Denk, T.~I.; Borsos, Z.; Engel, J.; Verzetti, M.; Caillon, A.; Huang, Q.; Jansen, A.; Roberts, A.; Tagliasacchi, M.; et~al. 2023.
\newblock Musiclm: Generating music from text.
\newblock \emph{arXiv preprint arXiv:2301.11325}.

\bibitem[{Arnab et~al.(2021)Arnab, Dehghani, Heigold, Sun, Lu{\v{c}}i{\'c}, and Schmid}]{arnab2021vivit}
Arnab, A.; Dehghani, M.; Heigold, G.; Sun, C.; Lu{\v{c}}i{\'c}, M.; and Schmid, C. 2021.
\newblock Vivit: A video vision transformer.
\newblock In \emph{Proceedings of the IEEE/CVF international conference on computer vision}, 6836--6846.

\bibitem[{Brunner et~al.(2018)Brunner, Konrad, Wang, and Wattenhofer}]{brunner2018midivae}
Brunner, G.; Konrad, A.; Wang, Y.; and Wattenhofer, R. 2018.
\newblock MIDI-VAE: Modeling Dynamics and Instrumentation of Music with Applications to Style Transfer.
\newblock arXiv:1809.07600.

\bibitem[{Copet et~al.(2024)Copet, Kreuk, Gat, Remez, Kant, Synnaeve, Adi, and D{\'e}fossez}]{copet2024simple}
Copet, J.; Kreuk, F.; Gat, I.; Remez, T.; Kant, D.; Synnaeve, G.; Adi, Y.; and D{\'e}fossez, A. 2024.
\newblock Simple and controllable music generation.
\newblock \emph{Advances in Neural Information Processing Systems}, 36.

\bibitem[{Corner(2002)}]{Corner_2002}
Corner, J. 2002.
\newblock Sounds real: music and documentary.
\newblock \emph{Popular Music}, 21(3): 357–366.

\bibitem[{Di et~al.(2021)Di, Jiang, Liu, Wang, Zhu, He, Liu, and Yan}]{di2021video}
Di, S.; Jiang, Z.; Liu, S.; Wang, Z.; Zhu, L.; He, Z.; Liu, H.; and Yan, S. 2021.
\newblock Video background music generation with controllable music transformer.
\newblock In \emph{Proceedings of the 29th ACM International Conference on Multimedia}, 2037--2045.

\bibitem[{Dong et~al.(2018)Dong, Hsiao, Yang, and Yang}]{Dong_Hsiao_Yang_Yang_2018}
Dong, H.-W.; Hsiao, W.-Y.; Yang, L.-C.; and Yang, Y.-H. 2018.
\newblock MuseGAN: Multi-track Sequential Generative Adversarial Networks for Symbolic Music Generation and Accompaniment.
\newblock \emph{Proceedings of the AAAI Conference on Artificial Intelligence}, 32(1).

\bibitem[{Dosovitskiy et~al.(2021)Dosovitskiy, Beyer, Kolesnikov, Weissenborn, Zhai, Unterthiner, Dehghani, Minderer, Heigold, Gelly, Uszkoreit, and Houlsby}]{dosovitskiy2021image}
Dosovitskiy, A.; Beyer, L.; Kolesnikov, A.; Weissenborn, D.; Zhai, X.; Unterthiner, T.; Dehghani, M.; Minderer, M.; Heigold, G.; Gelly, S.; Uszkoreit, J.; and Houlsby, N. 2021.
\newblock An Image is Worth 16x16 Words: Transformers for Image Recognition at Scale.
\newblock arXiv:2010.11929.

\bibitem[{Défossez et~al.(2022)Défossez, Copet, Synnaeve, and Adi}]{defossez2022highfi}
Défossez, A.; Copet, J.; Synnaeve, G.; and Adi, Y. 2022.
\newblock High Fidelity Neural Audio Compression.
\newblock \emph{arXiv preprint arXiv:2210.13438}.

\bibitem[{Evans et~al.(2024)Evans, Carr, Taylor, Hawley, and Pons}]{evans2024fast}
Evans, Z.; Carr, C.; Taylor, J.; Hawley, S.~H.; and Pons, J. 2024.
\newblock Fast Timing-Conditioned Latent Audio Diffusion.
\newblock arXiv:2402.04825.

\bibitem[{Forsgren and Martiros(2022)}]{Forsgren_Martiros_2022}
Forsgren, S.; and Martiros, H. 2022.
\newblock {Riffusion - Stable diffusion for real-time music generation}.

\bibitem[{Gan et~al.(2020)Gan, Huang, Chen, Tenenbaum, and Torralba}]{gan2020foley}
Gan, C.; Huang, D.; Chen, P.; Tenenbaum, J.~B.; and Torralba, A. 2020.
\newblock Foley music: Learning to generate music from videos.
\newblock In \emph{Computer Vision--ECCV 2020: 16th European Conference, Glasgow, UK, August 23--28, 2020, Proceedings, Part XI 16}, 758--775. Springer.

\bibitem[{Gemmeke et~al.(2017)Gemmeke, Ellis, Freedman, Jansen, Lawrence, Moore, Plakal, and Ritter}]{gemmeke2017audio}
Gemmeke, J.~F.; Ellis, D.~P.; Freedman, D.; Jansen, A.; Lawrence, W.; Moore, R.~C.; Plakal, M.; and Ritter, M. 2017.
\newblock Audio set: An ontology and human-labeled dataset for audio events.
\newblock In \emph{2017 IEEE international conference on acoustics, speech and signal processing (ICASSP)}, 776--780. IEEE.

\bibitem[{Huang et~al.(2018)Huang, Vaswani, Uszkoreit, Shazeer, Simon, Hawthorne, Dai, Hoffman, Dinculescu, and Eck}]{huang2018music}
Huang, C.-Z.~A.; Vaswani, A.; Uszkoreit, J.; Shazeer, N.; Simon, I.; Hawthorne, C.; Dai, A.~M.; Hoffman, M.~D.; Dinculescu, M.; and Eck, D. 2018.
\newblock Music Transformer.
\newblock arXiv:1809.04281.

\bibitem[{Huang et~al.(2022)Huang, Jansen, Lee, Ganti, Li, and Ellis}]{huang2022mulan}
Huang, Q.; Jansen, A.; Lee, J.; Ganti, R.; Li, J.~Y.; and Ellis, D. P.~W. 2022.
\newblock MuLan: A Joint Embedding of Music Audio and Natural Language.
\newblock arXiv:2208.12415.

\bibitem[{Hussain et~al.(2023)Hussain, Liu, Sun, and Shan}]{hussain2023m}
Hussain, A.~S.; Liu, S.; Sun, C.; and Shan, Y. 2023.
\newblock M$^2$UGen: Multi-modal Music Understanding and Generation with the Power of Large Language Models.
\newblock \emph{arXiv preprint arXiv:2311.11255}.

\bibitem[{Ji, Luo, and Yang(2020)}]{ji2020comprehensivesurveydeepmusic}
Ji, S.; Luo, J.; and Yang, X. 2020.
\newblock A Comprehensive Survey on Deep Music Generation: Multi-level Representations, Algorithms, Evaluations, and Future Directions.
\newblock arXiv:2011.06801.

\bibitem[{Kang, Poria, and Herremans(2024)}]{kang2024video2music}
Kang, J.; Poria, S.; and Herremans, D. 2024.
\newblock Video2Music: Suitable music generation from videos using an Affective Multimodal Transformer model.
\newblock \emph{Expert Systems with Applications}, 123640.

\bibitem[{Li et~al.(2022)Li, Li, Xiong, and Hoi}]{li2022blip}
Li, J.; Li, D.; Xiong, C.; and Hoi, S. 2022.
\newblock Blip: Bootstrapping language-image pre-training for unified vision-language understanding and generation.
\newblock In \emph{International conference on machine learning}, 12888--12900. PMLR.

\bibitem[{Li et~al.(2021)Li, Yang, Ross, and Kanazawa}]{Li_2021_ICCV}
Li, R.; Yang, S.; Ross, D.~A.; and Kanazawa, A. 2021.
\newblock AI Choreographer: Music Conditioned 3D Dance Generation With AIST++.
\newblock In \emph{Proceedings of the IEEE/CVF International Conference on Computer Vision (ICCV)}, 13401--13412.

\bibitem[{Li et~al.(2024)Li, Qin, Zheng, Jin, and Liu}]{Li_2024_CVPR}
Li, S.; Qin, Y.; Zheng, M.; Jin, X.; and Liu, Y. 2024.
\newblock Diff-BGM: A Diffusion Model for Video Background Music Generation.
\newblock In \emph{Proceedings of the IEEE/CVF Conference on Computer Vision and Pattern Recognition (CVPR)}, 27348--27357.

\bibitem[{Liu(1985)}]{liu_aesthetic_1985}
Liu, M. B.-R. 1985.
\newblock Aesthetic {Principles} in {Chinese} {Music}.
\newblock \emph{The World of Music}, 27(1): 19--32.
\newblock Publisher: [Florian Noetzel GmbH Verlag, VWB - Verlag für Wissenschaft und Bildung, Schott Music GmbH \& Co. KG, Bärenreiter].

\bibitem[{Radford et~al.(2021)Radford, Kim, Hallacy, Ramesh, Goh, Agarwal, Sastry, Askell, Mishkin, Clark, Krueger, and Sutskever}]{radford2021learningtransferablevisualmodels}
Radford, A.; Kim, J.~W.; Hallacy, C.; Ramesh, A.; Goh, G.; Agarwal, S.; Sastry, G.; Askell, A.; Mishkin, P.; Clark, J.; Krueger, G.; and Sutskever, I. 2021.
\newblock Learning Transferable Visual Models From Natural Language Supervision.
\newblock arXiv:2103.00020.

\bibitem[{Raffel et~al.(2020)Raffel, Shazeer, Roberts, Lee, Narang, Matena, Zhou, Li, and Liu}]{JMLR:v21:20-074}
Raffel, C.; Shazeer, N.; Roberts, A.; Lee, K.; Narang, S.; Matena, M.; Zhou, Y.; Li, W.; and Liu, P.~J. 2020.
\newblock Exploring the Limits of Transfer Learning with a Unified Text-to-Text Transformer.
\newblock \emph{Journal of Machine Learning Research}, 21(140): 1--67.

\bibitem[{Roberts et~al.(2018)Roberts, Engel, Raffel, Hawthorne, and Eck}]{roberts2018musicvae}
Roberts, A.; Engel, J.; Raffel, C.; Hawthorne, C.; and Eck, D. 2018.
\newblock A Hierarchical Latent Vector Model for Learning Long-Term Structure in Music.
\newblock In Dy, J.; and Krause, A., eds., \emph{Proceedings of the 35th International Conference on Machine Learning}, volume~80 of \emph{Proceedings of Machine Learning Research}, 4364--4373. PMLR.

\bibitem[{Su et~al.(2023)Su, Li, Huang, Kuzmin, Lee, Donahue, Sha, Jansen, Wang, Verzetti et~al.}]{su2023v2meow}
Su, K.; Li, J.~Y.; Huang, Q.; Kuzmin, D.; Lee, J.; Donahue, C.; Sha, F.; Jansen, A.; Wang, Y.; Verzetti, M.; et~al. 2023.
\newblock V2Meow: Meowing to the Visual Beat via Music Generation.
\newblock \emph{arXiv preprint arXiv:2305.06594}.

\bibitem[{Su, Liu, and Shlizerman(2020)}]{su2020audeo}
Su, K.; Liu, X.; and Shlizerman, E. 2020.
\newblock Audeo: Audio generation for a silent performance video.
\newblock \emph{Advances in Neural Information Processing Systems}, 33.

\bibitem[{Surís et~al.(2022)Surís, Vondrick, Russell, and Salamon}]{D2022It}
Surís, D.; Vondrick, C.; Russell, B.~C.; and Salamon, J. 2022.
\newblock It's Time for Artistic Correspondence in Music and Video.
\newblock \emph{2022 IEEE/CVF Conference on Computer Vision and Pattern Recognition (CVPR)}, 10554--10564.

\bibitem[{Tang et~al.(2024)Tang, Yang, Zhu, Zeng, and Bansal}]{tang2024any}
Tang, Z.; Yang, Z.; Zhu, C.; Zeng, M.; and Bansal, M. 2024.
\newblock Any-to-any generation via composable diffusion.
\newblock \emph{Advances in Neural Information Processing Systems}, 36.

\bibitem[{Vaswani et~al.(2017)Vaswani, Shazeer, Parmar, Uszkoreit, Jones, Gomez, Kaiser, and Polosukhin}]{vaswani2017attention}
Vaswani, A.; Shazeer, N.; Parmar, N.; Uszkoreit, J.; Jones, L.; Gomez, A.~N.; Kaiser, {\L}.; and Polosukhin, I. 2017.
\newblock Attention is all you need.
\newblock \emph{Advances in neural information processing systems}, 30.

\bibitem[{Wu et~al.(2024)Wu, Fei, Qu, Ji, and Chua}]{wu2024nextgptanytoanymultimodalllm}
Wu, S.; Fei, H.; Qu, L.; Ji, W.; and Chua, T.-S. 2024.
\newblock NExT-GPT: Any-to-Any Multimodal LLM.
\newblock arXiv:2309.05519.

\bibitem[{Xu et~al.(2021)Xu, Ghosh, Huang, Okhonko, Aghajanyan, Metze, Zettlemoyer, and Feichtenhofer}]{xu2021videoclip}
Xu, H.; Ghosh, G.; Huang, P.-Y.; Okhonko, D.; Aghajanyan, A.; Metze, F.; Zettlemoyer, L.; and Feichtenhofer, C. 2021.
\newblock Videoclip: Contrastive pre-training for zero-shot video-text understanding.
\newblock \emph{arXiv preprint arXiv:2109.14084}.

\bibitem[{Yu et~al.(2023)Yu, Wang, Chen, Sun, and Qiao}]{yu2023longtermrhythmicvideosoundtracker}
Yu, J.; Wang, Y.; Chen, X.; Sun, X.; and Qiao, Y. 2023.
\newblock Long-Term Rhythmic Video Soundtracker.
\newblock arXiv:2305.01319.

\bibitem[{Zeghidour et~al.(2021)Zeghidour, Luebs, Omran, Skoglund, and Tagliasacchi}]{zeghidour2021soundstream}
Zeghidour, N.; Luebs, A.; Omran, A.; Skoglund, J.; and Tagliasacchi, M. 2021.
\newblock SoundStream: An End-to-End Neural Audio Codec.
\newblock arXiv:2107.03312.

\bibitem[{Zhu et~al.(2022)Zhu, Olszewski, Wu, Achlioptas, Chai, Yan, and Tulyakov}]{zhu2022quantized}
Zhu, Y.; Olszewski, K.; Wu, Y.; Achlioptas, P.; Chai, M.; Yan, Y.; and Tulyakov, S. 2022.
\newblock Quantized GAN for Complex Music Generation from Dance Videos.
\newblock arXiv:2204.00604.

\bibitem[{Zhu et~al.(2023)Zhu, Wu, Olszewski, Ren, Tulyakov, and Yan}]{zhu2023discrete}
Zhu, Y.; Wu, Y.; Olszewski, K.; Ren, J.; Tulyakov, S.; and Yan, Y. 2023.
\newblock Discrete Contrastive Diffusion for Cross-Modal Music and Image Generation.
\newblock arXiv:2206.07771.

\bibitem[{Zhuo et~al.(2023)Zhuo, Wang, Wang, Liao, Bao, Peng, Han, Zhang, Fang, and Liu}]{zhuo2023video}
Zhuo, L.; Wang, Z.; Wang, B.; Liao, Y.; Bao, C.; Peng, S.; Han, S.; Zhang, A.; Fang, F.; and Liu, S. 2023.
\newblock Video background music generation: Dataset, method and evaluation.
\newblock In \emph{Proceedings of the IEEE/CVF International Conference on Computer Vision}, 15637--15647.

\end{thebibliography}

\end{document}